\documentclass[fleqn,twoside]{article}
\usepackage{espcrc2}
\usepackage{graphicx}
\usepackage{mciteplus}
\usepackage{amssymb,amsmath}
\usepackage[rflt]{floatflt}
\providecommand{\href}[2]{#2}

\newcommand{\qs}{{Q_\mathrm{s}}}

\newcommand{\lqcd}{\Lambda_{\mathrm{QCD}}}
\newcommand{\as}{{\alpha_{\mathrm{s}}}}

\newcommand{\gev}{\textrm{ GeV}}
\newcommand{\fm}{\textrm{ fm}}

\newcommand{\npart}{{N_\textrm{part}}}

\newcommand{\ud}{\, \mathrm{d}}
\newcommand{\fig}{Fig.~}

\title{RHIC data and small $x$ physics}

\author{T. Lappi\address{Department of Physics
 P.O. Box 35, 40014 University of Jyv\"askyl\"a, Finland
}
\address{Institut de Physique Th\'eorique,
B\^at. 774, CEA/DSM/Saclay, 91191 Gif-sur-Yvette Cedex, France}%
\thanks{
Supported by the Academy of Finland, contract 126604.}
}
\begin{document}

\begin{abstract}
This is a short review of some RHIC results that have been most important for the 
small $x$ physics community. We discuss saturation effects in deuteron-gold collisions, 
particle production in gold-gold collisions and some effects of the large ``glasma'' 
field configurations in the early stages of the collision.
\vspace{1pc}
\end{abstract}

\maketitle

\section{RHIC experiments}

The Relativistic Heavy Ion collider (RHIC) at Brookhaven National Laboratory (BNL)
(see e.g. Ref.~\cite{Nagle:2002wj})
has been colliding gold and lighter nuclei at an energy of
$\sqrt{s} = 200 A \gev$ since 2000. The accelerator also delivers polarized 
proton beams at $\sqrt{s} = 500 \gev$, used to study spin physics.
The main goal of the experimental program is the
discovery and study of the Quark Gluon Plasma (see 
e.g. Refs.~\cite{Adams:2005dq,*Adcox:2004mh,*Arsene:2004fa,*Back:2004je} 
for overall reviews of the experimental results).

There have been four major RHIC experiments, the two smaller ones now 
decommissioned and the two larger ones being upgraded and taking more data:
\begin{description}
\item[PHOBOS] is a small detector, mostly of silicon. It has a
huge acceptance in pseudorapidity $\eta$ and azimuthal angle $\phi$, but mostly without $p_T$
measurement or particle identification.
\item[BRAHMS] is also a small detector, consisting of 
2 moveable spectrometer arms that give it a large $\eta$ coverage with 
particle identification, but a small acceptance.
\item[PHENIX] is a large detector optimized more for electromagnetic signals 
(photons, leptons) with a good acceptance but not full azimuthal coverage
($\Delta \phi=\pi$ in two spectrometer arms).
\item[STAR] resembles most closely a typical high energy experiment, being 
built around a large TPC covering the full azimuthal angle. It is more optimized
towards hadronic and global observables.
\end{description}

The near term future of the BNL program is to continue taking data and improving the
PHENIX and STAR detectors in parallel to the LHC heavy ion program. In the longer term
there are plans to build an Electron-Ion Collider (EIC) to collide electrons with high
energy protons and nuclei~\cite{Deshpande:2005wd}. At CERN the LHC will, in addition to 
the proton beams, have lead-lead (and eventually other species) collisions at 
$\sqrt{s} = 5500 A \gev$. There is one detector (ALICE) dedicated to heavy ions and smaller
heavy ion programs in the CMS and ATLAS experiments.

\begin{figure*}
\centerline{\includegraphics[width=0.9\textwidth,clip=true]{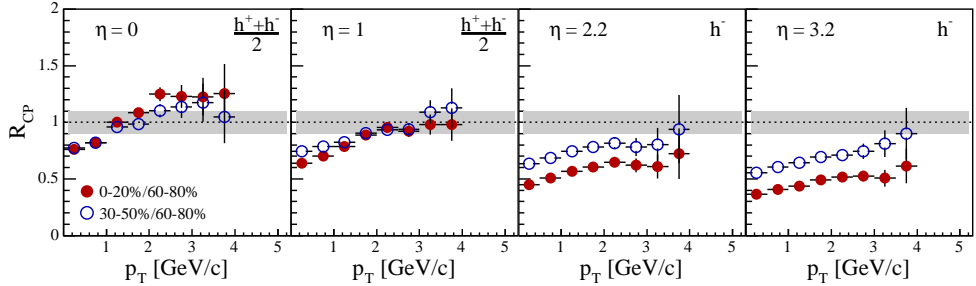}}
\vspace{-0.6cm}
\caption{BRAHMS results on suppression of charged hadron production in the deuteron
fragmentation region~\cite{Arsene:2004ux}. The plot shows $R_{CP}$ (see text) for charged hadron spectra
between two central bins (0-20\% and 30-50\%) and peripheral collisions (60-80\% most central).
}
\label{fig:brahms}
\end{figure*}

An important tool for understanding the geometry of the collision goes by the name
of Glauber modeling. Experimental data are usually presented divided into centrality
bins determined mostly by the total charged multiplicity. 
A Glauber model treats 
the nucleus as a collection of loosely bound nucleons distributed 
according to the charge density distribution.
These nucleons then scatter independently with the experimentally
measured proton-proton cross section. A Glauber model the associates each impact parameter
with a certain number of \emph{participant nucleons} $N_\textrm{part}$ and 
\emph{binary collisions} $N_\textrm{bin}$. One thus obtains an estimate for the 
typical impact parameter corresponding to the centrality bins. While the assumption 
of independent collisions is not a very realistic one at high energies, Glauber 
modeling has turned out to be a convenient way to parametrize the geometrical aspects
of the collision; experimental data are usually expressed in terms of $N_\textrm{part}$
and $N_\textrm{bin}$.
Experimental results
 for nuclear modifications to the spectra of produced particles
are often presented as ratios 
(such as $R_{AA}$, ratio of 
nucleus-nucleus to proton-proton; $R_{dAu}$, deuteron-gold to proton-proton
and $R_{CP}$, central to peripheral)
corrected by the geometrical factor deduced from a Glauber calculation.
These ratios are defined 
in such a way that in the case of independent scatterings of the nucleons in the nucleus
they should be equal to one. 

We can understand the time evolution of the collision process in different stages:
\begin{enumerate}
\item 
The initial condition at $\tau = 0$ depends on the properties
of the nuclear wavefunction at small $x$. 
\item  
Thermal and chemical equilibration.
\item 
The Quark Gluon Plasma, surviving for some fermis
around $\tau_0 \lesssim\tau \lesssim 10 \fm$. 
If the system reaches local thermal equilibrium, finite temperature 
field theory and relativistic hydrodynamics
can be used to describe its behaviour.
\item 
Finally, for $\tau  \gtrsim 10 \fm$ the system hadronises and decouples.
\end{enumerate}
In the remainder of this talk we will concentrate on the physics of the first stage,
the initial condition.

\begin{figure*}
\centerline{\includegraphics[width=0.9\textwidth,clip=true]{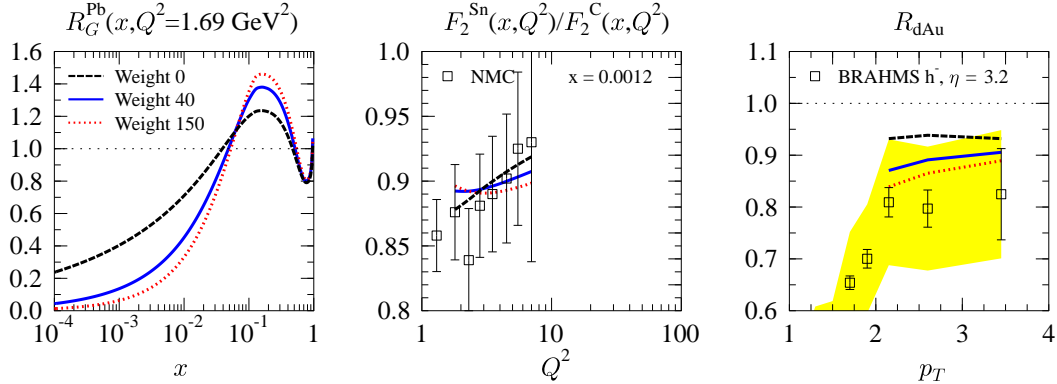}}
\vspace{-0.6cm}
\caption{DGLAP parametrizations of nuclear pdf's from Ref.~\cite{Eskola:2008ca}. 
The left panel shows the  suppression of the nuclear gluon distribution
required to
simultaneously fit the nuclear DIS data from NMC (center panel) and 
charged hadron production data from BRAHMS (right panel) and PHENIX data.}
\label{fig:eps}
\end{figure*}

\section{Color glass and glasma}

The central rapidity region in high energy collisions originates from the interaction 
of the ``slow'' small $x$ degrees of freedom, predominantly gluons,
 in the wavefunctions of the incoming hadrons or nuclei.
At large energies these gluons form a dense system 
that is characterized by a transverse momentum scame $\qs$, \emph{the saturation scale}.
The degrees of freedom with $p_T \lesssim \qs$ are fully nonlinear Yang-Mills 
fields with large field
strength $A_\mu \sim 1/g$ and occupation numbers $\sim 1/\as$, 
they can therefore be understood
as classical fields radiated from the large $x$ partons. Note that while
this description is inherently 
nonperturbative, it is still based on weak coupling,
because the classical approximation requires 
$1/\as$ to be large and therefore $\qs \gg \lqcd$.
A ``pocket formula''
for estimating the energy and nuclear dependence of the saturation scale is 
$\qs^2 \sim A^{1/3} x^{-0.3}$: nonlinear high gluon density effects are enhanced by going to 
small $x$ and large nuclei. Ideally one would like to study the physics of the CGC at the Electron Ion 
Collider~\cite{Deshpande:2005wd}, but already based on fits to HERA data and simple nuclear geometry
we have a relatively good idea of the magnitude of $\qs$ at RHIC energies~\cite{Kowalski:2007rw}.
The CGC is a systematic effective theory (effective because the large $x$ part of the wavefunction is 
integrated out) formulation of these degrees of freedom. The term \emph{glasma}~\cite{Lappi:2006fp} refers to 
the coherent, classical field configuration resulting from the collision of two such objects CGC.

\section{Deuteron-nucleus collisions}
\label{sec:pA}

A theoretically more controlled way to study the small-$x$ nuclear wavefunction is
using proton-nucleus (in practice deuteron-nucleus) collisions, especially at forward rapidity,
i.e. in the fragmentation region of the proton. In this case final state interactions in  the
plasma phase are expected to be absent, and the relevant degrees of freedom in the proton or 
deuteron are the relatively well understood large $x$ parton distributions. 

\begin{figure}
\centerline{\includegraphics[width=0.35\textwidth,angle=270,clip=true]{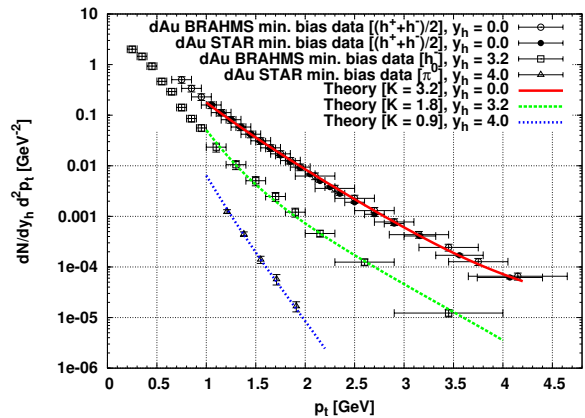}}
\vspace{-0.6cm}
\caption{
CGC calculation of charged hadron production~\cite{Dumitru:2005kb}.
} \label{fig:pAspectra}
\end{figure}

The effects of parton saturation are particularly visible in the BRAHMS data
(Fig.~\ref{fig:brahms}) for charged hadron 
production, which shows a significant depletion towards the deuteron fragmentation region.
Although these data DGLAP-based \cite{Eskola:2008ca} analysis can reproduce this data,
see  \fig\ref{fig:eps}, this requires that the gluon distribution at small $x$ is suppressed
by a factor of $\sim$10 compared to the proton. This is a clear signal of the breakdown of the DGLAP 
picture and the importance of saturation effects. The suppression pattern at large rapidities 
was predicted from the CGC~\cite{Kharzeev:2003wz} and later more detailed calculations
 have quite successfully reproduced it, see Fig.~\ref{fig:pAspectra}.

\section{Particle production and geometry}

\begin{figure}
\centerline{\includegraphics[width=0.35\textwidth]{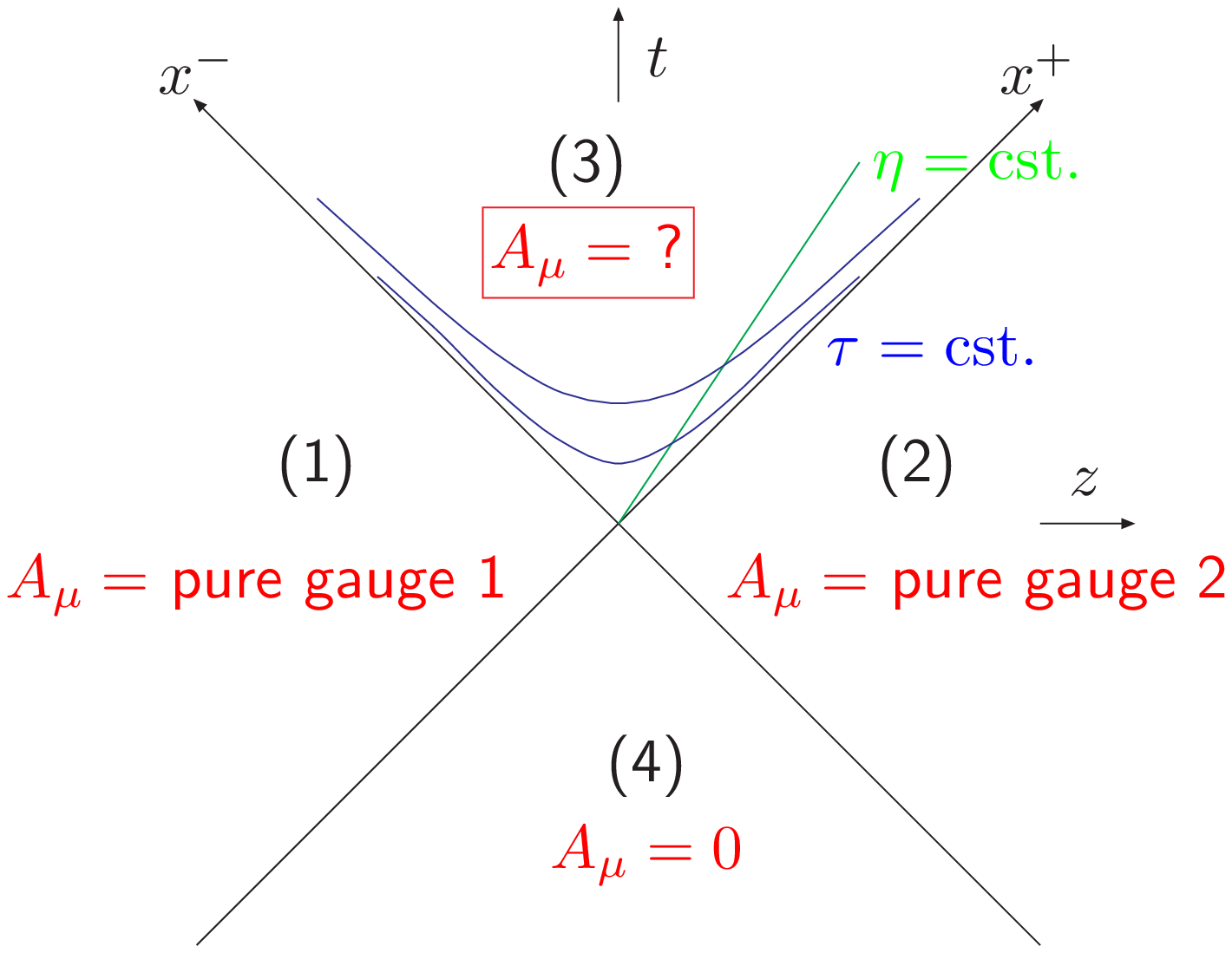}}
\vspace{-0.6cm}
\caption{Classical field configurations.}
\label{fig:spacet}
\end{figure}
In order to compute particle production in the Glasma one is presented
with the following situation \cite{Kovner:1995ts}.
The valence-like degrees of freedom of the two nuclei are represented
by two classical color currents that are, because of their large longitudinal momenta ($p^\pm$)
well localized on the light cone (in the variables conjugate to $p^\pm$, namely $x^\mp$):
 $J^\pm \sim \delta(x^\mp)$.
These then generate the classical field that one wants to find. Working in 
light cone gauge  the field in the region
of spacetime causally connected to only one of the nuclei
(regions (1) and (2) in Fig.~\ref{fig:spacet}) is a transverse pure gauge, 
independently  for each of the two nuclei.
These pure gauge fields then give the initial condition on the future 
light cone ($\tau = \sqrt{2 x^+ x^-} = 0$) for the nontrivial
gauge field after the collision (region (3) in Fig.~\ref{fig:spacet}).
The field inside the future light cone can then be computed
either numerically~\cite{Krasnitz:2001qu,*Lappi:2003bi,*Krasnitz:2003jw} or
analytically in different approximations (see e.g.~\cite{Blaizot:2008yb} for recent work).
The obtained result is then averaged over the configurations of the sources $J^\mu$ with a distribution
$W_y[J^\mu]$ that includes the nonperturbative knowledge of the large $x$ degrees of freedom.
The resulting fields are then decomposed into Fourier modes to get the gluon spectrum.
This is the method that we will refer to as Classical Yang-Mills (CYM) calculations.

In the limit when either one or both of the color sources are dilute (the ``pp'' and ``pA'' cases), the
CYM calculation can be done analytically and reduces to a factorized form in terms of a convolution of
unintegrated parton distributions that can include saturation effects.
Although this approach (often known as ``KLN'' after the authors \mbox{of~\cite{Kharzeev:2000ph,*Kharzeev:2001gp})}
is not is not derived for the collision of two dense systems, it has advantage of offering
more analytical insight and making it easier to incorporate large-$x$ ingredients into the calculation.

\begin{figure}
\centerline{\includegraphics[width=0.4\textwidth]{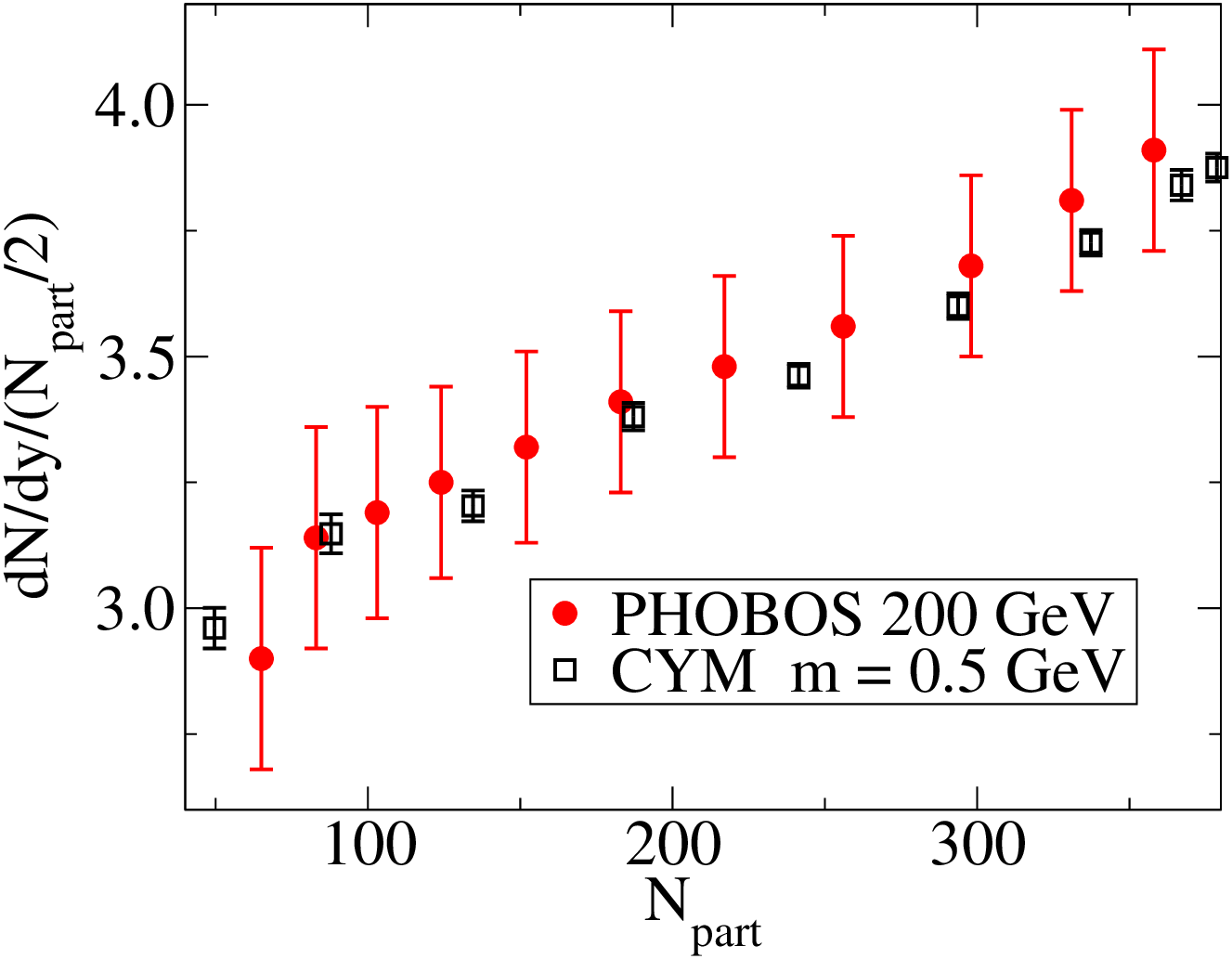}}
\vspace{-0.6cm}
\caption{
Centrality dependence of the multiplicity from a CYM calculation~\cite{Lappi:2006xc}.
} \label{fig:centrality}
\end{figure}

The CYM calculations~\cite{Krasnitz:2001qu} of gluon production paint a fairly consistent picture
of gluon production at RHIC. The estimated value $\qs \approx 1.2 \gev$ from HERA 
data~\cite{Kowalski:2007rw} (corresponding to the MV model parameter 
$g^2 \mu \approx 2.1 \gev$~\cite{Lappi:2007ku}) leads to $\frac{\ud N}{\ud y} \approx 1100$ 
gluons in the initial stage. Assuming a rapid thermalization and nearly ideal hydrodynamical evolution
this is consistent with the observed $\sim 700$ charged ($\sim 1100$ total) particles produced in a unit of 
rapidity in central collisions. 

Of the more detailed geometrical aspects of the initial condition 
the basic features (such as the closeness to $\npart$ scaling) of the centrality dependence of particle
multiplicities are mostly straightforward consequences of the proportionality
of the multiplicity to $\qs^2$; they are successfully reproduced by 
both KLN and CYM calculations~\cite{Kharzeev:2000ph,Lappi:2006xc,Drescher:2007ax}, 
see \fig\ref{fig:centrality}.

A striking signal of collective behavior of the matter
produced at RHIC is elliptic flow; where the initial \emph{spatial} anisotropy 
in a noncentral collision is transformed to an anisotropic distribution of produced 
particles in \emph{momentum} space through collective flow caused by the pressure gradients.
Comparing hydrodynamical calculations with measurements 
is a way to address fundamental properties of the medium, such the equation of state 
or viscosity,
but this requires understanding of the initial conditions.
These initial conditions
have traditionally been phenomenologically parametrized in terms of the number of nucleons
participating in the collision. The CGC provides a method to compute them
from first principles, and has changed the interpretation of the experiments in terms of 
the viscosity~\cite{Lappi:2006xc,Hirano:2005xf,*Drescher:2006ca}.

\begin{figure*}
\centerline{\includegraphics[height=150pt,clip=true]{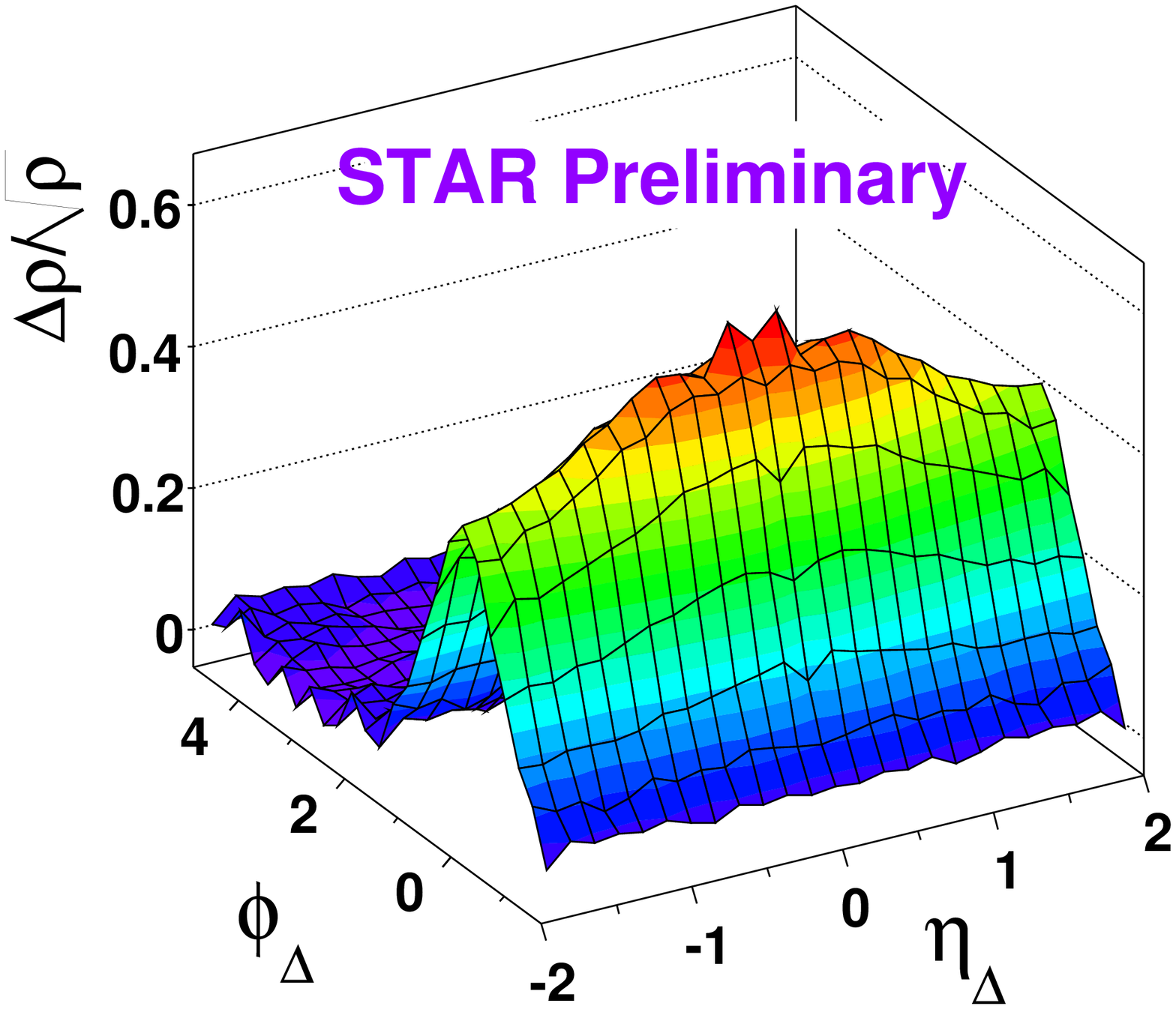}
\hfill
\includegraphics[height=150pt,clip=true]{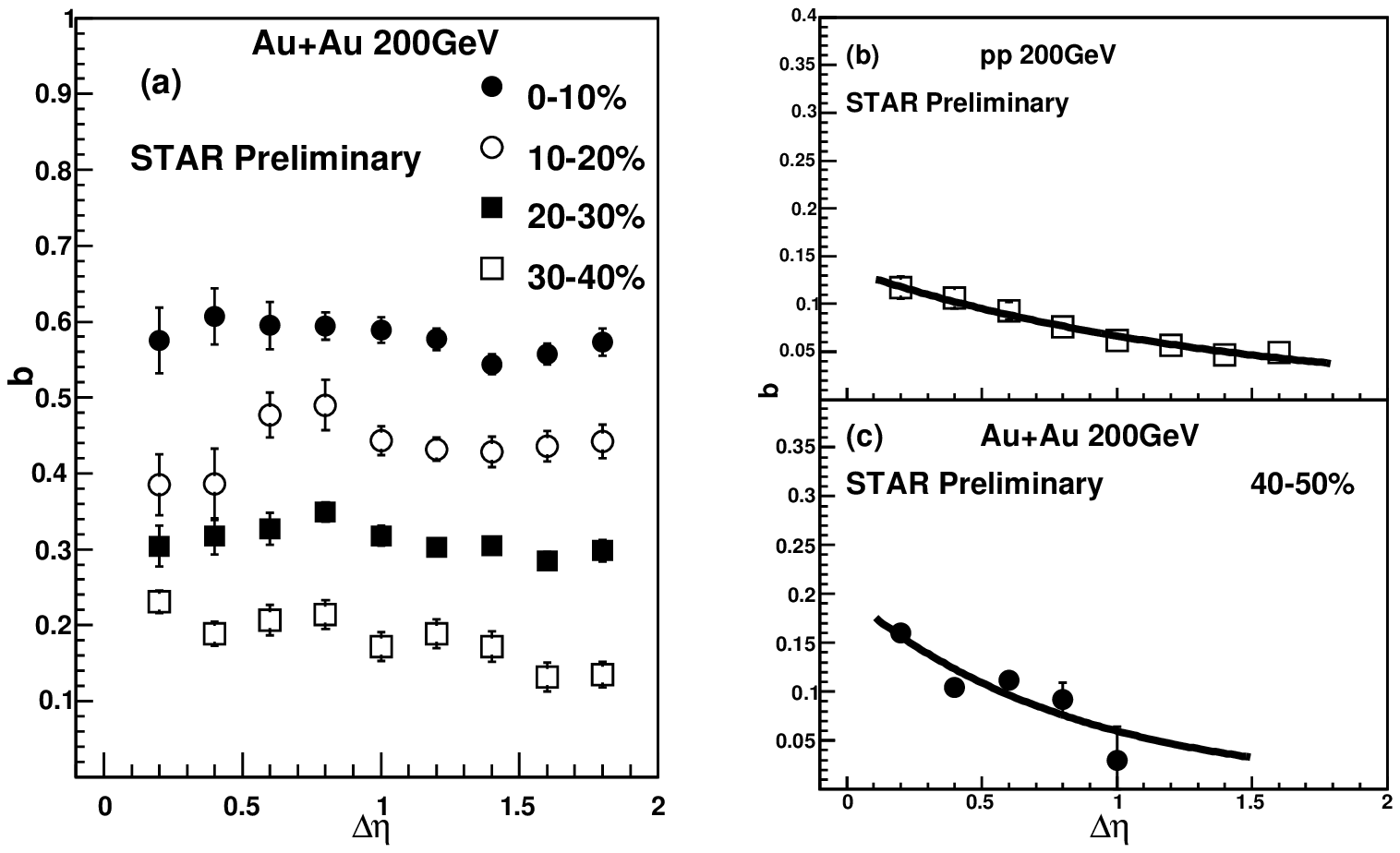}
}
\vspace{-0.6cm}
\caption{Left: Two particle correlation measurement from STAR, showing the ``ridge'' 
structure elongated in pseudorapidity.
Right: Long range correlations in multiplicity:
$b = \frac{\langle N_F  N_B \rangle - \langle N_F \rangle \langle N_B \rangle}
{\langle N_F^2 \rangle - \langle N_F \rangle^2 },$ where $F$ and $B$ are pseudorapidity
bins separated by $\Delta \eta$.
}
\label{fig:ridgelongrange}
\end{figure*}

An example of a more exclusive probe of the medium is the $J/\Psi$. It has been called 
the ``thermometer of the QGP'', because it the state is expected from lattice calculations
to melt at a temperature close to the QCD phase transition. The production mechanism
of the  $J/\Psi$ from perturbative QCD is not very well understood, and since the charm quark mass
is close to the typical values $\qs$, it can be expected to be very sensitive to
saturation physics~\cite{Kharzeev:2008nw}.

\begin{figure}
\centerline{\includegraphics[width=0.4\textwidth]{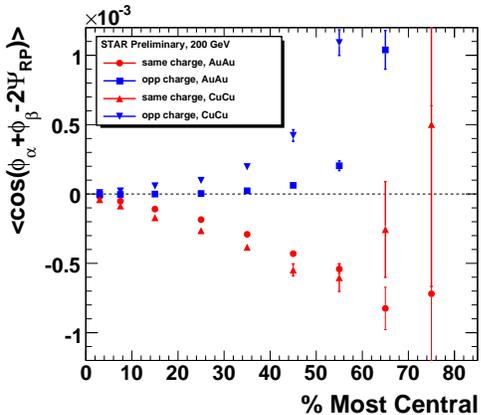}}
\vspace{-0.6cm}
\caption{
Parity violating correlation
between reaction plane and momentum~\cite{Voloshin:2008jx}. }
\label{fig:cpviol}
\end{figure}

\section{Glasma physics}
\label{sec:glasma}

Correlations over large distances in rapidity can, 
by causality, only originate from the earliest
times in the collision process, and are thus sensitive to the 
properties of the glasma phase of 
the collision. Examples of such phenomena are the elongated
``ridge'' structure seen in two particle correlations 
 and  long range correlations in multiplicity (see \fig\ref{fig:ridgelongrange})
 \cite{Putschke:2007mi,*Daugherity:2008su,*Srivastava:2007ei}.
The boost invariant nature of the Glasma fields
naturally leads to this kind of a correlation, and deviations from it should
be calculable from the high energy evolution governing the rapidity 
dependence~\cite{Dumitru:2008wn,*Gelis:2008sz}.

Another remarkable phenomenon that is possible in  the Glasma field
configuration is the generation of a large Chern-Simons charge 
and thus parity violation~\cite{Kharzeev:2001ev,*Kharzeev:2007jp}
due to the nonperturbatively large field configurations. Through the so called
``chiral magnetic effect''  this can manifest itself in a 
parity violating correlation between the electric dipole moment (or momentum anisotropy 
between negative and positive charges; a vector) and the reaction plane
(the positive charges of the ions generate a magnetic field perpendicular
to the reaction plane; a pseudovector). There are some preliminary indications 
in the data of a nonzero value for such an observable~\cite{Voloshin:2008jx}.

In conclusion, experiments at RHIC have clearly produced 
a strongly interacting deconfined form of matter, but the full understanding of the
experimental results requires an understanding of the small $x$ physics giving the initial 
conditions of the collision process. This is the domain of strongly nonliear gluon fields at
high energy, best understood in terms gluon saturation and the Color Glass Condensate
effective theory. This framework provides a way to calculate properties of the initial condition
from first principles, relate them to DIS measurements, and understand some of the observables
that directly probe the initial condition. With the much greater collision energy of the LHC these 
effects are likely to become even more prominent.

\bibliographystyle{h-physrev4mod2M}
\bibliography{spires}

\end{document}